\documentclass[12pt]{article}
\usepackage{cite}
\usepackage{times} \usepackage{amssymb,amsbsy}
\usepackage{pifont}\usepackage{amsmath}
\usepackage{graphics}
\newcommand\email[4]{#1@#2.#3.#4}

\def\R{{\mathbf{R}}}

\def\eq#1{(\ref{#1})}

\renewcommand\hat{\widehat}
\newcommand\ie{\emph{i.e.}}

\newcommand\pa{\partial}
\def\map{\longmapsto}
\def\ghat{\widehat{g}}
\def\k{\mathbf{k}}
\def\x{\mathbf{x}}
\def\y{\mathbf{y}}
\begin{document}
\title{Scale invariance and superfluid turbulence}
\author{ 
Siddhartha Sen  
\thanks{\email{siddhartha.sen}{tcd}{ie}, \email{tcss}{iacs}{res}{in}} \\ 
\small CRANN, Trinity College Dublin, Dublin -- 2, Ireland
\\\small \&\\
\small  R.K. Mission Vivekananda University,
\small Belur-711202, West Bengal, India.
\and 
Koushik Ray \thanks{\email{koushik}{iacs}{res}{in}} \\
\small Department of Theoretical Physics,
\small  Indian Association for the Cultivation of Science \\
\small  Calcutta 700~032. India.
}
\maketitle
\vfil
\begin{abstract}
\noindent 
We construct a Schroedinger field theory 
invariant under local spatial 
scaling. It is shown to provide an effective theory of 
superfluid turbulence by
deriving, analytically,  the observed Kolmogorov $5/3$ law 
and to lead to a Biot-Savart interaction between the observed
filament excitations of the system as well.
\end{abstract}
\thispagestyle{empty}
\clearpage
\section{Introduction}
We describe a construction of an
$1+3$-dimensional Schroedinger field theory which is 
invariant under local scaling in the three spatial dimensions. 
This is effected by introducing  a gauge field and a spatial metric. 
The requirement of local scaling in three dimensions
allows for a Chern Simons term in the action but forbids a Maxwell term.  
The locally scale invariant action is unique in the sense 
that it contains all possible terms
having polynomial interaction among the Schroedinger field,
the gauge field and the metric.
Moreover, gauge invariance for this system is rather novel 
due to the presence of the metric. 

Historically, local scale invariance  was introduced 
\cite{weyl} in an attempt to
unifying the theories of gravitation and 
electromagnetism. Stipulating local scale invariance of 
the theory of General Relativity in four dimensions lead to
the extremely novel idea of introducing a gauge field with an additional term
in the action resembling  Maxwell's theory. 
Identifying this term as electromagnetism a unification of the
theories of gravitation 
and electromagnetism was deemed to be have been achieved through
purely geometric means. This approach was criticised, however,  
as being incompatible with the observed discrete spectra of atoms
\cite{einstein}. 
The idea was thus given up as a means to producing a unified field theory 
only to be revived later on with the local scaling of
lengths replaced by a local change of phase of a quantum wave function
\cite{London}. This construction is now known as ``gauge theory"
although it is no longer related to length scales. What is retained, however,
is the idea of introducing a gauge field in order to render a system 
invariant under a local symmetry. 

In this article we consider a spatially scale invariant generalization
of the Schroedinger field theory 
in the original sense of Weyl. Unlike the original approach, however,  
gauge variations, that is local changes of scale, 
are compensated for by a Ricci term rather than by a Maxwell term.
The theory constructed would be relevant for describing  any three-dimensional
non-relativistic quantum field theory which has scale invariance properties. 
A good testing ground for this effective theory is the 
phenomenon of turbulence in superfluid helium.

Turbulence in superfluid liquid Helium \cite{tsubota} 
exhibits Kolmogorov scaling \cite{srineevasan} and does not have
any associated discrete spectrum. Thus it could be a good testing 
ground for our effective theory.
The usual theoretical approaches to superfluid turbulence
uses the non-linear Gross-Pitaevski (GP) 
equation \cite{GP} where the nonlinearity reflects
the interaction between helium atoms in the field theory
description of the system taking a superfluid 
condensate into account. 

When energy is injected into the system, say by  
heating, excitations in the form of filaments appear. 
In this approach the observed filament excitations 
are understood with their location given by the zeros of the GP wave function. 
The filament excitations can also be modelled more directly with their 
dynamics described in analogy with interaction of wires carrying
currents obeying the Biot-Savart law \cite{schwarz}.

Distribution functions in superfluid turbulence are different from 
those arising for classical fluids. For instance, the velocity 
distribution function
is not Gaussian but has a power law tail \cite{kivo}.
We find that the unique locally scale invariant theory constructed here
contains the appropriate
degrees of freedom for describing superfluid turbulence
namely, a condensate and the filament excitations.
Indeed, we show that the theory
constructed predicts the Kolmogorov ${5}/{3}$ law observed in a certain
range of momenta of the quasi-particle excitations of the theory. 
While numerical studies of the GP equation yield similar results \cite{koba},
the present analysis is completely  analytic. 

The model presented can accommodate both filaments and
condensates which appear in all approaches used to describe
superfluid turbulence. The locus of the zeros of the condensate field gives
the location of the filaments, which are modelled as currents coupled to the
gauge field supported only on the filaments. 
An effective theory for the Schroedinger field is obtained by 
integrating out the gauge field from the 
theory. This produces the action for a GP-like system, while
the effective theory for the filaments is 
that of Wilson lines interacting through the Chern Simons term.
This leads
to a Biot-Savart type of interaction between the filaments.
This interaction can be used to estimate
the velocity of separation between a pair of filaments that collide. 
The result we obtain, valid for small times, 
is in agreement with the experimental results \cite{srineevasan}.  
Let us point out that the present approach examines the
consequences of the idea of local scale invariance but does not seek
to furnish a picture of the emergence of 
Kolmogorov scaling microscopically or, for instance, by
vortex tangles and Kelvin-wave turbulence caused by Kelvin waves on a 
single vortex. It provides an effective theory for a locally scale 
invariant system, superfluidity being an interesting example. 

The plan of the article is as follows. 
In section~\ref{s:action} we derive the unique scale invariant
action starting from the action of a free Schroedinger field. 
A solution to the equations of motion
for a special arrangement of filaments
is presented in section~\ref{s:soln}. In section~\ref{s:effective} we obtain
effective theories for the Schroedinger field and Wilson lines
before concluding in section~\ref{s:conclu}.
\section{Scale invariant action}
\label{s:action}
In this section we construct an action invariant under spatial scaling 
starting from the action for a free Schroedinger field. 
The gauge group is $\R^{\star}$, the group of non-zero reals, which is
non-compact.
The free Schroedinger equation, in operator form, 
allows Bose-Einstein condensation
and is thus an appropriate starting point for a theory of quantum turbulence.
First, the free system is made invariant under global scaling by introducing a
time-independent metric for the three spatial directions. It is then
made invariant under local scaling by introducing a gauge field.

The action of the Schroedinger field $\psi$
in $\R^1\times \R^3$, with the first factor designating 
time, $t$, and the second one corresponding to the spatial coordinates
$\mathbf{x}=(x^1,x^2,x^3)=(x,y,z)$ is 
\begin{equation} 
\label{action1}
\mathcal{S}(\psi,g) =  
i\int\psi^{\star}\pa_t\psi\sqrt{g}\,dt\,d^3x
-\frac{1}{2m}\int g^{ij}\pa_i\psi^{\star}\pa_j\psi \sqrt{g}\,dt\,d^3x,
\end{equation} 
where we have introduced a metric $g$ on $\R^3$ and $\pa_i$ denotes
the derivative with respect to $x^i$ and an asterisk designates complex
conjugation. 
The second term of the
action is invariant under the global scaling transformation of the field
$\psi$ and the metric
\begin{equation} 
\begin{gathered}
\psi \map e^{-\Lambda/4}\psi,\\
g_{ij}\map e^{\Lambda}g_{ij},
\end{gathered}
\end{equation} 
where $\Lambda$ is a constant. 
Let us note that the scale invariance could not be 
effected without the metric.
Moreover, as mentioned before, we do not impose scale invariance on the first
term involving temporal derivative of the Schroedinger field. 
We now promote this global scaling symmetry to a local symmetry by
allowing spatial dependence of $\Lambda$ \cite{colm} and
introducing a gauge field $A_i$ and define covariant derivatives of the
field $\psi$ and the metric $g$ as \cite{paddy}
\begin{equation} 
\begin{gathered}
\label{cov:der}
D_i\psi = \pa_i\psi -\alpha A_i\psi,\\
D_i g_{km} = \pa_ig_{km}+4\alpha A_ig_{km},
\end{gathered}
\end{equation}  
where $\alpha$ is a real parameter. It appears from \eq{cov:der} that the
parameter $\alpha$ may be dispensed with by a redefinition of the gauge
field. However, the sign of $\alpha$ is of import in obtaining field
configurations and will be fixed later. 
Then under the gauge transformation 
\begin{equation} 
\label{gauge}
\begin{gathered}
\psi \map e^{-\Lambda(\mathbf{x})/4}\psi,\\
g_{ij}\map e^{\Lambda(\mathbf{x})}g_{ij},\\
A_i\map A_i-\frac{1}{4\alpha}\pa_i\Lambda(\mathbf{x}),
\end{gathered}
\end{equation} 
with space-dependent $\Lambda$, the covariant derivatives of the scalar
field $\psi$ and the metric transform as
\begin{equation}
\label{cov:tr}
\begin{gathered}
D_i\psi \map e^{-\Lambda(\mathbf{x})/4}D_i\psi,\\
D_ig_{jk}\map e^{\Lambda(\mathbf{x})} D_ig_{jk}.
\end{gathered}
\end{equation} 
Hence replacing the derivatives
with respect to the spatial coordinates in the second term of \eq{action1}
by covariant derivatives we obtain  the action 
\begin{equation} 
\label{action}
\mathcal{S}(\psi,A,g) =  i\int\psi^{\star}\pa_t\psi\sqrt{g}\,dt\,d^3x-
\frac{1}{2m}\int
g^{ij}D_i\psi^{\star}D_j\psi \sqrt{g}\,dt\,d^3x,
\end{equation} 
which is invariant under the gauge transformations \eq{gauge}.
One can add one more gauge-invariant term to the above action 
involving the curvature and the gauge field \cite{iorio}. 
To this end let us define Christoffel symbols \cite{paddy} as
\begin{equation} 
\tilde{\Gamma}^i_{jk} = \frac{1}{2}g^{im}(D_jg_{mk}+D_kg_{mj}-D_mg_{jk}).
\end{equation} 
By \eq{cov:tr}, the Christoffel symbol is invariant under
the local scaling transformations \eq{gauge}.
Then the Ricci tensor ensuing from this Christoffel symbol defined as
\begin{equation}
\tilde{R}^i_{jkl} = 
\pa_l\tilde{\Gamma}^i_{jk} - \pa_k\tilde{\Gamma}^i_{jl}
+\tilde{\Gamma}^i_{ml}\tilde{\Gamma}^m_{jk}
-\tilde{\Gamma}^i_{mk}\tilde{\Gamma}^m_{jl}
\end{equation} 
is also invariant under the gauge transformation \eq{gauge}.
The resulting scalar curvature defined as
\begin{equation}
\tilde{R} = g^{jl}\tilde{R}^i_{jil} 
\end{equation} 
then transforms as $\tilde{R}\map e^{-\Lambda}\tilde{R}$ under \eq{gauge}.
Hence, 
\begin{equation} 
\label{ricciact}
\int|\psi|^2\tilde{R}\sqrt{g}\,dt\,d^3x
\end{equation} 
is invariant under the gauge transformation. It can be checked
that no other term involving curvature tensors or derivatives of $A$  or
their combinations can be made gauge invariant in this fashion to yield a
local polynomial action. In particular, 
the term $F^2_{ij}$ constructed from the gauge field $A_i$
is not scale invariant in three dimensions, nor can it be made gauge
invariant in a polynomial action. 

The Ricci scalar $\tilde{R}$ defined above can be related to the Ricci scalar
corresponding to the metric $g$ by expanding $\tilde{\Gamma}^I_{jk}$ using
\eq{cov:der} \cite{paddy}, resulting into  
\begin{equation}
\label{ricci} 
\begin{split}
\tilde{R} &= R + 8\alpha \nabla_iA^i+8\alpha^2A^2,\\
&= R + \frac{8\alpha}{\sqrt{g}}\pa_i(\sqrt{g}A^i)+8\alpha^2A^2,
\end{split}
\end{equation} 
where we used  $A^2=g^{ij}A_iA_j=A^iA_i$, $\nabla_i$ and $R$ denote,
respectively, the covariant
derivative with respect to $x^i$
and the scalar curvature corresponding to the metric $g$.

Putting \eq{ricci} in \eq{ricciact} and
adding  to \eq{action} along with a Chern Simons term for the gauge field
we obtain the unique spatially scale-invariant action 
in $1+3$ dimensions given by  
\begin{equation}
\label{actionfin}
\begin{split}
\mathcal{S}(\psi,A,g) &=  
\int\sqrt{g} dtd^3x \left(i\psi^{\star}\pa_t\psi 
 -\frac{1}{2m} g^{ij} (\pa_i\psi^{\star}\pa_j\psi 
- \alpha A_i\pa_j|\psi|^2+\alpha^2 A_iA_j|\psi|^2)\right) \\
&\;  +\beta\int dtd^3x |\psi|^2\left(\sqrt{g} R + 8\alpha\pa_i(\sqrt{g}A^i) +
8\alpha^2\sqrt{g} A^2\right)
+ \gamma\int dtd^3x\epsilon^{ijk}A_i\partial_j A_k,
\end{split}
\end{equation} 
where $\epsilon^{ijk}$ denotes the rank three antisymmetric tensor,
$\beta$ and $\gamma$ are real parameters.  
The Chern-Simons term being independent of the metric is
locally scale invariant. From the four-dimensional perspective, this term
is to be thought of as the unique potential term for the gauge field which
has local three-dimensional scale invariance. 
Furthermore, while it does not
contribute to the equations of motion, this term plays a crucial role,
as we shall see below, in determining the interaction between filaments
modelled by the gauge field. 
We now proceed to study the properties of this unique locally 
scale invariant three-dimensional system.
\section{Solution for special geometrical arrangement}
\label{s:soln}
Physical configurations are obtained as solutions to the Euler-Lagrange
equations ensuing from the action \eq{actionfin} by variation of the fields. 
The equation obtained  upon varying the metric $g$ is 
\begin{equation} 
\begin{split}
\label{eomg}
&-\frac{1}{2} g_{ij}\left(i\psi^{\star}\pa_t\psi - \frac{1}{2m}
(\pa_i\psi^{\star} \pa^i\psi-\alpha A^i\pa_i|\psi|^2 + \alpha^2 A^2 |\psi|^2)
+ \beta|\psi|^2\left(R+8\alpha\nabla_iA^i+8\alpha^2 A^2\right)
\right)\\
&-\frac{1}{2m}(\pa_i\psi^{\star}\pa_j\psi-\alpha A_i\pa_j|\psi|^2 
+\alpha^2 A_iA_j|\psi|^2) 
+ \beta|\psi|^2\left(R_{ij} +8\alpha\nabla_iA_j+ 8\alpha^2 A_iA_j\right) =0
\end{split}
\end{equation} 
Equations arising from the variations of the gauge field 
$A$ and the Schroedinger
field $\psi^{\star}$ are, respectively,
\begin{gather}
\label{eomA}
\pa_i|\psi|^2 = 2{\alpha}A_i|\psi|^2, \\
\label{eompsiwg}
i\sqrt{g}\pa_t\psi + \frac{1}{2m}\pa_i(\sqrt{g}
g^{ij}\pa_j\psi)+
\beta\sqrt{g}\left(R + \alpha (8-\frac{1}{2m\beta})\nabla_iA^i+ 
\alpha^2(8-\frac{1}{2m\beta})A^2\right)\psi=0,
\end{gather}
where we  assumed
\begin{equation}
16m\beta \neq 1. 
\end{equation} 
The Chern Simons term does not contribute to the equations of motion.

Let us focus on the equations of motion for stationary configurations, 
$\pa_t\psi=0$.
Assuming the Schroedinger field to be time-independent,
equations \eq{eomA} and  \eq{eomg} together lead to
\begin{equation}
\label{ricpsi}
R+8\alpha\nabla_iA^i+\alpha^2(8+\frac{1}{2m\beta})A^2=\frac{1}{2m\beta}
\frac{\pa_i\psi^{\star}\pa^i\psi}{|\psi|^2},
\end{equation} 
where in deriving this division by $|\psi|^2$ was used, so that the equation
is valid only for non-vanishing $|\psi|^2$.
For stationary configurations thus we need only to consider equations
\eq{eomA}, \eq{ricpsi} and \eq{eompsiwg} with the first term set to zero..

Equation \eq{eomA} is solved with 
\begin{equation}
\label{psiA}
|\psi|^2 = |\psi_0|^2\exp{\left(2\alpha\int\limits_C A_idx^i\right)},
\end{equation} 
where $\psi_0$ is a constant  
and $C$ the curve over which the line integral is evaluated. 

Filaments can be introduced in this model now by introducing currents
$J(\x)=J_{Ci}dx^i$ supported on a curve $C$, by delta functions,
$J_{Ci}=J^{(0)}_i\delta^{(3)}_{C}$, where $J^{(0)}_i$ is a constant.
We thus have
\begin{equation}
\int\limits_CA_i dx^i = 
\int\limits_{\R^3} A_i J^i(C)d^3x,
\end{equation} 
leading to
\begin{equation}
\label{psiJ}
|\psi|^2 = |\psi_0|^2\exp{\left(2\alpha J^{(0)i}\int\limits_{\R^3} 
A_i \delta^{(3)}_{(C)}d^3x\right)}.
\end{equation}  
We also relate the constant $\psi_0$ to the number of filaments $N$ as
$\psi_0=\sqrt{N/V}$, $V$ denoting the volume of the superfluid.
Choosing the constant $\alpha$ to be negative and $J^{(0)i}$ to be positive
by convention, the modulus of $\psi$ vanishes on the curve and is a non-zero
constant everywhere else, equal to $|\psi_0|^2$.
The constant wave function can be interpreted as representing
a Bose-Einstein condensate. 
The filaments then represent excitations of the system corresponding to 
injection of energy.
Thus the simple model has features which suggest that excitations can be 
described either as filaments or as zeros of the condensate. 
Moreover, the metric is $\eta_{ij}$ in the bulk of the condensate. 
In view of 
this in the next section we proceed to construct an effective action 
for the system in terms of the wave function by 
integrating out the gauge field and in
terms of the gauge field by integrating out 
the wave function.
\section{Effective theories}
\label{s:effective}
As mentioned in the introduction, two approaches for studying
superfluid turbulence are either using a GP equation 
of a  Bose-Einstein condensate or in terms
of interacting vortex filaments. 
In the previous section we found that the local
scale invariant theory allows for both of these 
configurations. We now proceed to 
construct an effective theory for the Schroedinger field 
by integrating out the gauge field from the action \eq{actionfin}.
As mentioned earlier, the scope of such a unique scale invariant
theory is rather vast. However, in view of the results of the previous 
section the metric pertaining to superfluid bulk is flat. 
Therefore, we set the metric to be the Euclidean one,
$g_{ij}=\eta_{ij}$ in the bulk of the condensate
and define $S(\psi, A)=S(\psi,A,\eta)$. 
As excitations in the superfluid background are filaments 
located at the zeroes of the Schroedinger field, 
the effective theory for the filaments is given by Wilson lines 
in the background of a Chern Simons theory. 
\subsection{The condensate and quasi-particle spectrum}
\label{sec:wt}
First let us integrate out the gauge field  from 
the action \eq{actionfin} with a flat metric to
obtain the effective action for the condensate $\psi$ defined by the path
integral 
\begin{equation}
\label{Seff}
e^{{iS_{\text{eff}}} (\psi)} = 
\frac{1}{\sqrt{\pi}}\int\mathcal{D}A e^{iS(\psi,A,\eta)}.
\end{equation} 
Setting $g_{ij}=\eta_{ij}$ in \eq{actionfin}, we obtain, up to boundary
terms 
\begin{equation} 
\label{seta}
\begin{split}
S(\psi, A,\eta) =
i\int\psi^{\star}\pa_t\psi &-\frac{1}{2m}\int\pa_i\psi^{\star}\pa_i\psi 
-\frac{\ghat}{4}
\int\left(\pa_i\log |\psi|^2\right)^2|\psi|^2 \\ &\,\,+
\ghat\int\left(A_i-\frac{1}{2}\pa_i\log |\psi|^2\right)^2|\psi|^2+
\gamma\int\epsilon^{ijk}{A}_i\pa_j {A}_k.
\end{split}
\end{equation} 
where we defined $\ghat = 8\alpha\beta-\frac{1}{2m}$ and suppressed the
measure $dtd^3x$ in the integrals.
We now redefine the gauge field with a shift, namely,
\begin{equation}
\tilde{A}_i=A_i-\frac{1}{2}\pa_i\log |\psi|^2.
\end{equation} 
Then in the Chern-Simons term 
\begin{equation} 
\int\epsilon^{ijk}A_i\pa_j A_k = 
\int\epsilon^{ijk}\tilde{A}_i\pa_j \tilde{A}_k,
\end{equation} 
up to boundary terms.
Integrating out with respect to the new filed $\tilde{A}$ we
obtain the effective action
\begin{equation} 
\label{spsi}
\begin{split}
S_{\text{eff}}(\psi) =
i\int\psi^{\star}\pa_t\psi &-\frac{1}{2m}\int\pa_i\psi^{\star}\pa_i\psi 
-\frac{\ghat}{4}
\int\left(\pa_i\log |\psi|^2\right)^2|\psi|^2 
+\gamma\int\epsilon^{ijk}\tilde{A}_i\pa_j \tilde{A}_k + \Gamma.
\end{split}
\end{equation} 
where the effective potential 
\begin{equation} 
\Gamma = -\frac{1}{2}\int\limits_0^{L}\frac{d\xi}{\xi}\int d^3x 
e^{-\xi\ \ghat|\psi|^2}.
\end{equation}
Expanding the exponential and performing the integration with respect to
$\xi$, the effective potential becomes
\begin{equation}
\label{gsum}
\Gamma =
-\frac{1}{2}\sum\limits_{n=1}^{\infty}\frac{(-1)^n}{n!}
\frac{(\ghat L)^n}{n}\int |\psi|^{2n},
\end{equation} 
where we have neglected an infinite constant term ensuing from the unit term
in the exponential.

Let us now consider the quasi-particle spectrum of this theory \cite{wlt,rawk}. Considering
stationary configurations we expand
$\psi$ in Fourier modes 
\begin{equation}
\label{modex}
\psi(x) = \frac{1}{\sqrt{V}}\sum_{\mathbf{k}} a_{\k}e^{i\k\cdot\x}, 
\quad\psi^{\star}(x) = 
\frac{1}{\sqrt{V}}\sum_{\mathbf{k}} a_{\k}^{\dagger}e^{-i\k\cdot\x}, 
\end{equation} 
where $a_{\k}^{\dagger}$ and $a_{\k}$ are, respectively, creation and
annihilation operators for the bosonic modes satisfying the commutation
relation
\begin{equation} 
\label{commrel}
[a^{\dagger}_{\k},a_{\k'}] = \delta_{\k\k'}. 
\end{equation} 
The sum is over all momentum
modes. For each momentum mode we define a number operator
$\hat{n}(k)=a_{\k}^{\dagger}a_{\k}$, depending only on the magnitude of the
momentum, thanks to the rotational symmetry. The states diagonalizing these
number operators satisfy 
\begin{equation} 
\hat{n}(k)|n(\k)\rangle = n(k)|n(\k)\rangle, \quad
a_{\k}|n(\k)\rangle = \sqrt{n(k)}|n(\k)-1\rangle, \quad
a_{\k}^{\dagger}|n(\k)\rangle = \sqrt{n(k)+1}|n(\k)+1\rangle.
\end{equation}  
For the zero momentum mode we
also assume the existence of a state $|\psi_0\rangle=|n(0)\rangle$ with  
\begin{equation}
a_0 |\psi_0\rangle = a^{\dagger}_0 |\psi_0\rangle = \sqrt{N}|\psi_0\rangle,
\end{equation} 
where we denoted $n(0)=N$ and assumed $N$ to be sufficiently large so that
$\sqrt{N}\sim\sqrt{N+1}$. This state corresponds to the condensate over which
the non-zero modes are taken to be fluctuations. 
Substituting \eq{modex}  in \eq{gsum} we obtain
\begin{equation}
\Gamma =
-{\frac{1}{2}}\!\!
\sum_{\genfrac{}{}{0pt}{}{\k_1',\k_2'\cdots ,\k_n'}{\k_1,\k_2\cdots ,\k_n}}
\sum\limits_{n=1}^{\infty}\frac{(-1)^n}{n!}
\frac{g^n}{n} 
a^{\dagger}_{\k_1'} a^{\dagger}_{\k_2'} \cdots a^{\dagger}_{\k_n'}
a_{\k_1} a_{\k_2} \cdots a_{\k_n}
\delta(\k_1'+\k_2'+\cdots +\k_n'-\k_1-\k_2-\cdots -\k_n),
\end{equation} 
where we denoted $g=\ghat L/V$.
So far we have not fixed the parameters. We now assume that $g=1/N$.
Then in $\Gamma$ the quadratic terms  
$a^{\dagger}_{\k}a_{\k}$, 
$a_{-\k}a_{\k}$ and $a^{\dagger}_{-\k}a^{\dagger}_{\k}$
arise with $N^{n-1}$ in the $n$-th term, while all other terms are lower
order in $N$. The effective potential becomes
\begin{equation} 
\begin{split}
\Gamma =
-\frac{1}{2}\sum_{\k}
\sum\limits_{n=1}^{\infty}\frac{(-1)^n}{n!}
\frac{g}{n} 
\left(n^2 a^{\dagger}_{\k} a_{\k} 
+\binom{n}{2} a_{-\k}a_{\k}
+\binom{n}{2} a^{\dagger}_{-\k} a^{\dagger}_{\k} \right) 
+ \mathcal{O}(1/N)\times\text{quartic terms}.
\end{split}
\end{equation} 
The coefficients of the quadratic terms are determined by the number of
ways of satisfying the momentum conservation constraint
\begin{equation} 
\k_1'+\k_2'+\cdots +\k_n'=\k_1+\k_2+\cdots +\k_n.
\end{equation} 
For example, the term $a^{\dagger}_{\k}a_{\k}$ is obtained as one chooses any
one $\k'$ as well as any single $\k$ to be non-zero, which can be chosen in
$n\times n$ ways. The term $a_{-\k}a_{\k}$ is obtained by choosing all $\k'$
to be zero and two of the $n$ $\k$'s to be non-zero. The third term is
obtained similarly. 

Now, along with the kinetic term, the Hamiltonian reads, upon performing the
sum over $n$,
\begin{equation}
H = \sum_{\k\neq 0} \left(2\ell_1 a^{\dagger}_{\k} a_{\k} 
- \ell_2 (a_{-\k} a_{\k}+a^{\dagger}_{-\k} a^{\dagger}_{\k} )\right),
\end{equation} 
where we defined 
\begin{equation} 
\ell_1  = \frac{1}{2}\left(\frac{k^2}{2m}+\frac{g}{2e}\right),\quad
\ell_2= \frac{g}{2e}(\frac{e}{2}-1).
\end{equation} 
Let us note that the third term involving the derivative of $\ln |\psi|^2$
comes with the coupling constant $\ghat = g V/L$, which can be ignored
compared with $g$. 
In order to obtain the quasi-particle spectrum we need to diagonalize the
Hamiltonian. To this end we change basis as
\begin{gather}
a_{\k}= u \alpha_{\k}+v\alpha^{\dagger}_{-\k} \\
a^{\dagger}_{\k}= u \alpha^{\dagger}_{\k}+v\alpha_{-\k}, 
\end{gather} 
where $u$ and $v$ are taken to be real parameters. 
Requiring the commutations relations 
\begin{equation}
[\alpha^{\dagger}_{\k},\alpha_{\k'}]=\delta_{\k\k'},
\end{equation} 
in addition to \eq{commrel}
for any momentum $\k$, we obtain the constraint $u^2-v^2=1$, so that
the two bases are related by a Bogoliubov
transformation 
\begin{gather}
a_{\k}= \alpha_{\k}\cosh\theta + \alpha^{\dagger}_{-\k}\sinh\theta \\
a^{\dagger}_{\k}= \alpha^{\dagger}_{\k}\cosh\theta +\alpha_{-\k}\sinh\theta, 
\end{gather} 
where $\theta$ is a real parameter. 
Then expressing the Hamiltonian in terms of the new oscillators $\alpha$ and
demanding that the off-diagonal terms vanish, we obtain a relation among
$\theta$, $\ell_1$ and $\ell_2$, namely
\begin{gather}
\ell_1\cosh 2\theta-\ell_2\sinh 2\theta = \frac{1}{2}\epsilon(k) \\
\ell_1\sinh 2\theta -\ell_2 \cosh 2\theta=0,
\end{gather} 
where $\epsilon(k)$ is the dispersion depending on the magnitude $k$ of the
quasi-particle momentum $\k$ due to the rotational symmetry. 
Solving for the hyperbolic functions in terms of $\ell_1$, $\ell_2$ and
$\epsilon(\k)$, and using the identity 
$\cosh^22\theta-\sinh^22\theta=1$, yields an expression of $\epsilon(k)$ in
terms of $\ell_1$ and $\ell_2$, which in turn relates it to the $g$,
\begin{equation}
\label{ek}
\begin{split}
\epsilon(k)  &= 2 (\ell_1^2-\ell_2^2)^{1/2}\\ 
&= \left(\left(\frac{k^2}{2m}+\frac{g}{2e}\right)^2 
-\left(\frac{g}{e}\right)^2\left(\frac{e}{2}-1\right)^2\right)^{1/2},
\end{split}
\end{equation} 
and the Hamiltonian is 
\begin{equation}
H = \sum_{\k\neq 0}\epsilon(k)\ n(k),
\end{equation} 
where $n(k)= \alpha^{\dagger}_{\k} \alpha_{\k}$ is the occupation number of 
the quasi-particle state with energy $\epsilon$ and momentum $\k$, depending
on the magnitude of $\k$ again thanks to the rotational symmetry. 
\subsubsection*{Kolmogorov Scaling}
From our locally scale invariant model we have seen that in order
to describe the superfluid state a quasiparticle with  energy that scales linearly with momentum in a certain range of momentum values emerges. Unlike the
standard Bogoliubov quasiparticle result the second sound value is
not fixed by the theory but has to be taken from experiment but the
linear relationship between energy and momentum is present in both approaches in the small momentum region. This result was obtained,
as in the Bogoliubov approach, by putting in details of the
superfluid helium state in terms of a condensate. For this calculation
to be valid  the interaction between quasiparticles must be small. This is essential for the idea of quasiparticle to be useful.  These two features,
both present in our effective model, allow 
us to use the method of weak wave turbulence to determine the
turbulent properties of superfluid helium.  Essentially this means determining  the  way energy for momentum $k$ scale with momentum and to check if the scaling exponent of energy calculated agrees with the observed Kolmogorov 
exponent. 

 We have already found the scaling exponent for quasiparticles. 
We now need to find if the occupation number for momentum $k$ scales
with $k$ and to determine this exponent. The key calculation to do this,
in weak turbulence, involves setting up a Boltzmann type of equation for the
occupation number of quasiparticles with a given momentum and checkiing to
see if it has a time independent scaling solution \ie a solution where the occupation number scales with the momentum  exists.  The procedure outlined is a 
standard step of weak wave turbulence.  The  calculation for a quartic interaction term has been done and we can simply use the known results
to write down scaling exponent for occupation number for our case.  Once
this exponent is determined the Kolmogorov exponent is fixed.
 
For our quasiparticle system with a weak  quartic interaction term
 the Boltzmann time evolution equation for the quasiparticle excitation number
of momentum $\k$ is obtained from the system Hamiltonian.
Time independent  solutions to this evolution equation with a scaling law behaviour have energy \cite{falkovich,rawk,sanyal,balk}
\begin{equation}
\label{kolmo}
\begin{split}
E(k) &= n(k) \epsilon(k) \\ &\sim k^{-{\gamma}/{3}},
\end{split}
\end{equation}
where the exponent $\gamma$ is expressed as
$\gamma=\frac{3d+2 \beta}{\alpha}-4$ in  terms of the spatial dimension
$d$ and the exponents of scaling of the coefficient of the
quartic term and the energy dispersion, namely
\begin{gather}
T(k) \sim k^{\beta}\\
\epsilon(k)  \sim  k^{\alpha}.
\end{gather}
We have so far discussed the terms quadratic in the raising and lowering
operators in the Hamiltonian. The coefficient of the
 quartic term, which goes as $1/N$ in the
large $N$ limit that we are considering, is independent  of $\k$, leading to
$\beta=0$. 
As can be seen from \eq{ek} if the momenta are in the  range
\begin{equation}
\label{momres}
\begin{split}
\frac{k^2}{2m}< \frac{g}{e}\\
\frac{k^2}{2m}\left(\frac{k^2}{2m} +\frac{g}{e}\right) &>
\frac{g}{e}(e-\frac{e^2}{4}-\frac{3}{4})\\ 
&= 0.12 \left(\frac{g}{e}\right)^2,
\end{split}
\end{equation} 
then the dispersion is linear  in momentum and thus gives  $\alpha=1$.
Hence, $\gamma=5$, leading, according to \eq{kolmo}, to the Kolmogorov
scaling law, $E(k)\sim k^{-5/3}$, within this range of momentum. 
Thus, in an appropriate range of momentum we obtain linear dispersion
relation and thus weak turbulence and Kolmogorov scaling law from the four
wave resonance. 
\subsection{Filaments}
In section~\ref{s:soln} we obtained
field configurations with currents along curves describing filaments. 
Correlation between two filaments is then
understood as the correlation between current 
supported on a pair of curves, say, $C_1$ and $C_2$, as we shall discuss in
this section. The modulus of the 
condensate $\psi$ vanishes on the filaments and assumes a non-zero
constant value outside the filaments, as described by \eq{psiJ}. 
Adding source terms to the action
for the filaments $C_a$, $a=1,2, \cdots$,
and setting $\psi$ to zero the action becomes 
\begin{equation}
S_{\text{fil}} =  
\gamma\int\limits_{\R^3} d^3x\epsilon^{ijk}{A}_i\pa_j {A}_k
+\sum_a\int\limits_{\R^3} d^3x A_iJ_{C_a}^{i}.
\end{equation} 
Introducing the observables
$W_a=\exp\left( i\int d^3x A_i J^i(\x) \right)$, the correlator
$\langle W_1 W_2\rangle$ is obtained by 
integrating out the gauge field as
\begin{equation}
\begin{split}
\label{w1w2}
\langle W_1 W_2\rangle &= 
\exp \left({\frac{i}{2\gamma} \int\ dt\int d^3x\ \int 
d^3y J^i_{C_1}(\x)G_{ij}(\x-\y) J^j_{C_2}(\y)}\right),
\end{split}
\end{equation} 
where $G_{ij}(\x-\y)$ denotes the Green's function associated to the 
Chern-Simons term, given by
\begin{equation}
L_{ij} G_{jk}(\x-\y)=\delta_{ik}\ \delta^{(3)}(\x-\y),
\end{equation} 
where we defined, for compactness of notation, the operator
$L_{ij}=-\epsilon^{ijk}\pa_k$.
Operating with $L$ from the left on both sides we obtain
\begin{equation}
(L^2)_{ij}G_{jk}=L_{ik}\ \delta^{(3)}(\x-\y),
\end{equation} 
so that the Green's function is given by
\begin{equation}
\label{glinv}
G_{ij}=L_{kj}(L^{-2})_{ki}\ \delta^{(3)}(\x-\y).
\end{equation} 
Now using $(L^2)_{ij}= \pa_i\pa_j-\delta_{ij}\pa^2$, we have
\begin{equation} 
(L^2)_{ij}\frac{1}{|\x-\y|} =\pa_i\pa_j\frac{1}{|\x-\y|}
-\delta_{ij}\delta({|\x-\y|}),
\end{equation} 
leading to
\begin{equation}
(L^{-2})_{ij}\ \delta^{(3)}(\x-\y) = (L^{-2})_{ik}\pa_k\pa_j\frac{1}{|\x-\y|}
- \delta_{ij}\frac{1}{|\x-\y|}.
\end{equation} 
Using this formula in \eq{glinv} we obtain
\begin{equation}
G_{ij}= - L_{jk}\pa_k\pa_i\frac{1}{|\x-\y|}-L_{ij}\frac{1}{|\x-\y|}. 
\end{equation} 
The first term vanishes as $L_{jk}\pa_k\pa_i=\epsilon^{jkl}\pa_l\pa_k\pa_i=0$ 
leaving us with the expression for the Green's function
\begin{equation}
\begin{split}
G_{ij}(\x-\y) &=\epsilon^{ijk}\pa_k\frac{1}{|\x-\y|} \\
&=-\epsilon^{ijk} \frac{(\x-\y)_k}{|\x-\y|^3}.
\end{split}
\end{equation} 
Inserting this expression for the Green's function in \eq{w1w2}
the correlator assumes the form \cite{pol,witt,hans}
\begin{equation}
\label{ww}
\langle W_1W_2\rangle =
\exp\left(\frac{i}{2\gamma}\int dt\int_{C_1}ds_1\int_{C_2}ds_2
\;\epsilon_{ijk}\frac{(\x-\y)^k}{|\x-\y|^3} J^i(\x(s_1))J^j(\y(s_2))\right),
\end{equation} 
where we have introduced affine variables $s_1$ and $s_2$
parametrizing the curves $C_1$ and $C_2$ respectively.
The interaction energy for a pair of filaments is, then,
\begin{equation} 
\int\limits_{C_1}ds_1\ {\mathbf{B}(s_1)}\cdot{\mathbf{J}({\x(s_1))}},
\end{equation} 
where the field 
\begin{displaymath}
{\mathbf{B}(s_1)}=\frac{1}{2\gamma}\int\limits_{C_2} ds_2\ 
\frac{\mathbf{J}(\y(s_2))\times \big(\x(s_1)-\y(s_2)\big)}{|\x(s_1)-\y(s_2)|^{3}}
\end{displaymath}
is analogous to a magnetic field generated 
by a current $\mathbf{J}$. The equation for a current element of 
unit mass is therefore
\begin{equation} 
\frac {d\mathbf{u}}{dt} =\mathbf{u} \times \mathbf{B},
\end{equation} 
where $\mathbf{u}=d\x/dt$.
Recalling  that the current is a one-form supported on the filament,
$J(\x)=J_{Ci}dx^i$, 
we then derive the equation of a point on one single filament as
\begin{gather} 
\frac{d\x}{dt} \sim  \x \times \mathbf{v}, 
\end{gather} 
with  velocity
\begin{gather} 
\mathbf{v}= \int\frac{(\x-\y)}{|\x-\y|^3} \times d\mathbf{r},
\end{gather} 
We thus have a Biot-Savart type interaction between filaments.
Let us  emphasize again that the 
interaction term which leads to the dynamics of filament interactions
is not put in by hand in our approach but appears naturally
from the requirement of local spatial scale invariance. 
The velocity of separation
between two colliding filaments within a short time $\Delta t$ after the 
collision can be estimated from this. 
Moreover, the force $F$ between two small segments 
$d\x$ and  $d\y$ on the two curves
separated by a small distance $\mathbf{r}$ has magnitude
\begin{equation} 
F= \frac{1}{{r}^3}\ d\x\times d\y\cdot\mathbf{r},
\end{equation} 
where $r=|\mathbf{r}|$.
The impulse $K=F \Delta t$ within  a short interval of time 
$\Delta t$ after the collision is given as
\begin{equation} 
K\approx\frac{1}{{(\Delta v)^2}/{\Delta t}},
\end{equation} 
where $\Delta v$ is the separation
velocity of the filaments in this time.
Assuming $K$ to be a constant for this short time we conclude that
\begin{equation} 
(\Delta v)^2 \propto {1}/{\Delta t},
\end{equation} 
in agreement with experimental observations \cite{srineevasan}.
\section{Conclusions \& discussions}
\label{s:conclu}
In this article we have constructed a 
theory invariant under local spatial scaling
which can describe Bose-Einstein condensation.
Constructing such a theory is of theoretical interest 
as the scale invariance necessitates the introduction of a
gauge field and a metric with a Ricci term added to the action. 
Appearance of the Ricci scalar and the gauge field in
a specific combination is crucial for gauge invariance. 
This procedure has been called Ricci gauging \cite{iorio}.
A three-dimensional Chern Simons term for the gauge
field is also allowed. Since the scale invariance prohibits any other term
the action thus constructed is unique.

We have shown that the solution to the classical equations of motion are
solved with a flat metric yielding a configuration of the Schroedinger field
that can be interpreted as a condensate by virtue of its vanishing in the
core and being constant elsewhere. Indeed, the solution \eq{psiJ} is
reminiscent of the Madelung tranformation used in the study of the GP
equation. However, in here we study stationary configurations, rather than
the dynamics of the Madelung phase. The condensate is taken to be the
classical configurations whose fluctuations in the effective theory have been
studied. We demonstrate that under suitable approximation, as
elaborated in section~\ref{sec:wt}, our GP-like effective action \eq{spsi}, 
obtained by integrating out the gauge field, yields
Kolmogorov spectrum for excitations \eq{modex} over the classical
configuration, the condensate. There are instances \cite{naza}, for example,
Kelvin-wave turbulence
\footnote{We thank the anonymous referee for pointing out this article.}
in which the same dispersion law appears.
We should emphasize that the $5/3$ law is \emph{not generic} in this model.
It is only a restrictive 
range of momenta \eq{momres} that is commensurate with the Kolmogorov 
scaling. The range further entails a free parameter $g$, inherited from 
the action,  which can only be determined experimentally. 
In this respect the present model supplements the GP equation, which can but
describe superfluid turbulence only qualitatively. 
The scope of the model is thus rather broad, and
while it predicts a Kolmogorov-like dispersion only as a special case, it may
be used to study excitations with momenta beyond this range, perhaps to
study the emergence of Kolmogorov scaling. 
At this point let us stress again that the presented theory is by no means a
\emph{microscopic} theory. The degrees of freedom dealt with in here are
completely classical. However, this furnishes a unified description of the
two ways of looking at tubulence in terms of vortices and filaments. 

Introducing currents along curves corresponding to the heating of the
superfluid then makes the Schroedinger field into a condensate
vanishing at the location
of the filaments when the density of filaments is low and  with constant
modulus in the bulk. Such an identification of filament excitations as
the zeros the GP wave functions 
is a standard assumption but here the identification
is not a mathematical ansatz but follows from the result established that 
filament locations
are zeros of the Schroedinger condensate. It is thus a dynamical result of
the locally scale invariant theory constructed.

We considered the effective theory for interaction between filaments.
We only need to consider an action with a number of filaments
interacting through the Chern-Simons term thanks to the vanishing of the
condensate along the filaments. We show that this yields
a  Biot-Savart type of interaction between filaments described by currents by
evaluating the propagator of the Chern-Simons theory, which is the standard 
means of modelling interaction in a field theory.
This picture is compatible with modelling of filament dynamics 
along the flow created by all other vortex elements. In obtaining the
propagator of the Chern-Simons term, we fixed the condensate $\psi$ to its
classical value, which corresponds to vanishing of the corresponding terms of 
the action \eq{seta} in the bulk. However, the filaments here are not
envisaged as being the Goldstone mode of the broken translational symmetry.
But the distinction may not be experimentally perceptible at this stage. On
the other hand, our preference of the present picture stems from the fact
that the local gauge principle fixed the action uniquely.

We have also shown how the velocity of separation between two filaments 
after collision depended on the time of separation. This is done using 
an impulse  approximation between filaments based on the derived Biot-Savart 
interaction. The result obtained is in agreement with observations
\cite{srineevasan}.

Let us point out that time evolution breaks scale invariance  in the 
effective theory  constructed
as the standard kinetic energy term scales differently from the other terms
if the spatial and temporal coordinates are scaled similarly.
As filaments with length scales appear 
in turbulent superfluid flows this breakdown
is acceptable. On the other hand, 
it is well known that turbulent
flows that represent  far from equilibrium
dynamically generated stationary configurations \cite{falkovich}
can have  scale invariance
for its energy distribution spectrum even though
the starting space-time dynamics is not scale invariant.

We thus conclude that an effective theory based on Weyl's original idea of gauge invariance 
as local scale invariance is compatible with the existing descriptions
used to  understand of superfluid turbulence including
the interaction dynamics of excitations.
Local scale invariance leads to correctly identifying the degrees of freedom 
and  leads to the dynamics of the excitations
in the superfluid turbulent phase.
It is satisfying that the effective theory links filament locations,
postulated to be filament currents which couple to scale gauge fields,
with the zeros of the GP-like equation. The approach described
to construct locally scale invariant systems is also of theoretical interest
as it is a very general method for constructing locally scale invariant 
effective theories. 
\section{Acknowledgement}
SS acknowledges the hospitality of the Department of Theoretical
Physics, IACS, during the period this work was carried out.
KR thanks Pushan Majumdar and Krishnendu Sengupta for useful discussions.


\begin{thebibliography}{99}
\bibitem{weyl}
H.~Weyl, ``Gravitation and Electricity", Sitzungsber. Preuss. Akad. Berlin
(1918) 465. Collected in \cite{einstein}.
\bibitem{einstein}
L.~ O'Raifeartaigh, ``The Dawning of Gauge Theory", Princeton Series in
Physics, 1997.
\bibitem{London}
F.~London, ``Quantum-mechanical interpretation of Weyl's theory", Zeit. F.
Phys., {\bf 42}:375,1927. 
\bibitem{tsubota}
M.~Tsubota, ``Quantum Turbulence", J. Phys. Soc. Jpn. {\bf 77},
111006.(2008).
[arXiv:0806.2737(cond-mat)]
\bibitem{srineevasan}
M.~Paoletti {\it et al}, ``Velocity Statistics Distinguish Quantum Turbulence
from Classical Turbulence", Phys. Rev. Lett. {\bf 101}.154501(2008).
\bibitem{GP}
E.~Gross, ``Structure of a quantized vortex in boson systems", Il Nuovo
Cimento {\bf 20}.(1961).454.\\
L.~Pitaevskii,  ``Vortex Lines in an Imperfect Bose Gas", Soviet Physics
JETP-USSR (Woodbury, New York: American Institute of Physics) 
{\bf 13}.(1961).451.
\bibitem{schwarz}
K.~Schwarz, ``Three-dimensional vortex dynamics in superfluid
${}^4\text{He}$: Line-Line and line-boundary interactions, Phys. Rev. {\bf
B31}(1985)5782.
\bibitem{kivo}
D.~Kivotides \emph{et. al.}, 
``Velocity spectra of superfluid turbulence", Europhys. Lett.,
{\bf 57}.845 (2002).
\bibitem{koba}
M.~Kobayashi and M.~Tsubota,
``Kolmogorov Spectrum of Superfluid Turbulence: Numerical Analysis of the
Gross-Pitaevskii Equation with a Small-Scale Dissipation"  
Phys. Rev. Lett. {\bf 94}, 065302 (2005)
\bibitem{colm}
C.~Connaughton and S.~Sen, ``Local scale invariance and weak wave
Turbulence", (Unpublished).
\bibitem{paddy}
T.~Padmanabhan, ``Conformal invariance, gravity and massive gauge theories",
Class. Quantum Grav. {\bf 2}(1985)L105.
\bibitem{iorio}
  A.~Iorio, L.~O'Raifeartaigh, I.~Sachs {\it et al.},
  ``Weyl gauging and conformal invariance,''
  Nucl.\ Phys.\  {\bf B495}, 433 (1997).
  [hep-th/9607110].
\bibitem{wlt}
S.\ Musher, A. Rubenchik and V.\ Zakharov,
``Weak Langmuir turbulence"
Phys. Rep. {\bf 252} (1995) 177.
\bibitem{rawk}
M.~Rakowski and S.~Sen, ``Quantum kinetic equation in weak turbulence",
Phys.\ Rev.\ {\bf E 53}, 586–590 (1996). arXiv:cond-mat/9510107.
\bibitem{falkovich}
V.~Zakharov, V.~L'vov and G.~Falkovich, ``Kolmogorov spectra of turbulence",
Springer, 1992.
\bibitem{balk}
A.~M.~Balk,
``On the Kolmogorov–Zakharov spectra of weak turbulence",
Physica {\bf D 139} (2000) 137–157.
\bibitem{sanyal}
D.~Sanyal and S.~Sen, ``Quantum weak turbulence", 
Ann. \ Phys.\ {\bf 321} 1327 (2006). arXiv:cond-mat/0402395.
\bibitem{pol}
A.~Polyakov, ``Fermi-Bose Transmutations Induced by Gauge Fields", 
Mod.Phys.Lett. {\bf A3} (1988) 325
\bibitem{witt}
E.~Witten,``Quantum Field Theory and the Jones Polynomial",
Commun.Math.Phys. {\bf 121} (1989) 351
\bibitem{hans}
T. H. Hansson, A. Karlhede and  M. Rocek,
``On Wilson Loops In Abelian Chern-simons Theories"
Phys.Lett. {\bf B225} (1989) 92
\bibitem{naza}
V. S. L'vov and S. Nazarenko, 
``Spectrum of Kelvin-wave turbulence in superfluids", JETP Lett
(Pis'ma v ZhETF) {\bf 91} (2010) 464-470.
\end{thebibliography}
\end{document}